\documentclass[10pt,letterpaper]{article}
\usepackage[top=0.85in,footskip=0.75in]{geometry}
\counterwithin*{equation}{section}
\usepackage{amsmath,amssymb}

\usepackage{cite}

\usepackage{changepage}

\usepackage[utf8x]{inputenc}

\usepackage{textcomp,marvosym}

\usepackage{cite}

\usepackage{nameref,hyperref}

\usepackage{microtype}
\DisableLigatures[f]{encoding = *, family = * }

\usepackage{array}

\usepackage[aboveskip=1pt,labelfont=bf,labelsep=period,justification=raggedright,singlelinecheck=off]{caption}

\usepackage{svg}

\usepackage{cite}

\usepackage{lastpage,fancyhdr,graphicx}
\usepackage{epstopdf}
\begin{document}
\vspace*{0.2in}

\begin{flushleft}
{\Large
\textbf\newline{Superspreading of SARS-CoV-2 in the USA} 
}
\newline
\\
Calvin Pozderac,
Brian Skinner
\\
\bigskip
Department of Physics, Ohio State University, Columbus, Ohio 43210, USA
\\
\bigskip

\end{flushleft}

\section*{Abstract}
A number of epidemics, including the SARS-CoV-1 epidemic of 2002-2004, have been known to exhibit superspreading, in which a small fraction of infected individuals is responsible for the majority of new infections. The existence of superspreading implies a fat-tailed distribution of infectiousness (new secondary infections caused per day) among different individuals. Here, we present a simple method to estimate the variation in infectiousness by examining the variation in early-time growth rates of new cases among different subpopulations. We use this method to estimate the mean and variance in the infectiousness, $\beta$, for SARS-CoV-2 transmission during the early stages of the pandemic within the United States. We find that $\sigma_\beta/\mu_\beta \gtrsim 3.2$, where $\mu_\beta$ is the mean infectiousness and $\sigma_\beta$ its standard deviation, which implies pervasive superspreading. This result allows us to estimate that in the early stages of the pandemic in the USA, over 81\% of new cases were a result of the top 10\% of most infectious individuals.

\section*{Introduction}
The temporal growth of an epidemic is often characterized by either a time scale (such as the doubling time) \cite{doubletime,doubletime2} or by the reproduction rate $R_0$, which indicates the average number of new infections produced by each infected individual \cite{mathbio}. Estimates of $R_0$ for the current pandemic of SARS-CoV-2 range from 1.4 to 3.8 \cite{R_0,R_02,estimates,R0s}. Neither of these numbers, however, gives any information about the distribution of infectiousness among individuals --- i.e., whether new infections arise relatively uniformly from all infected individuals, or whether new infections are driven primarily by a small number of highly infectious individuals. The latter case is commonly referred to as ``superspreading", and different epidemics exhibit superspreading to different degrees. For example, during the outbreak of SARS CoV-1 in 2002-2004, over 80\% of cases were observed to result from the top 20\% most infectious individuals \cite{superspreadinghist,superspread_past}. Understanding the degree of superspreading in the current pandemic of SARS-CoV-2 is crucial for developing strategies to mitigate continued spread and informing an educated reopening procedure \cite{reopen, reopen2, vespignani2020modelling, weiner2020projections}.

Here we present a simple and direct method to understand how the infectiousness (also called the ``reproduction rate" of the disease) varies among infected individuals. At late times after the onset of an epidemic, the number of infected individuals is large, and consequently any statistical fluctuations in the growth rate are relatively small, so that the growth rate is well characterized by the mean infectiousness, $\mu_\beta$. However, at early times, when there are relatively few cases, the growth rate is stochastic and the degree of randomness depends on the variance in infectiousness, $\sigma_\beta^2$, between individuals (Fig 1a). By examining the variance in growth rate across subpopulations at these early times (Fig 1b), we are able to infer the variation in the distribution of infectiousness. In our analysis we divide the US cases into counties and observe how the variance in growth rate across them evolves as the number of cases increases.

\begin{figure}[htbp]
\centering
\includegraphics[scale = 0.75]{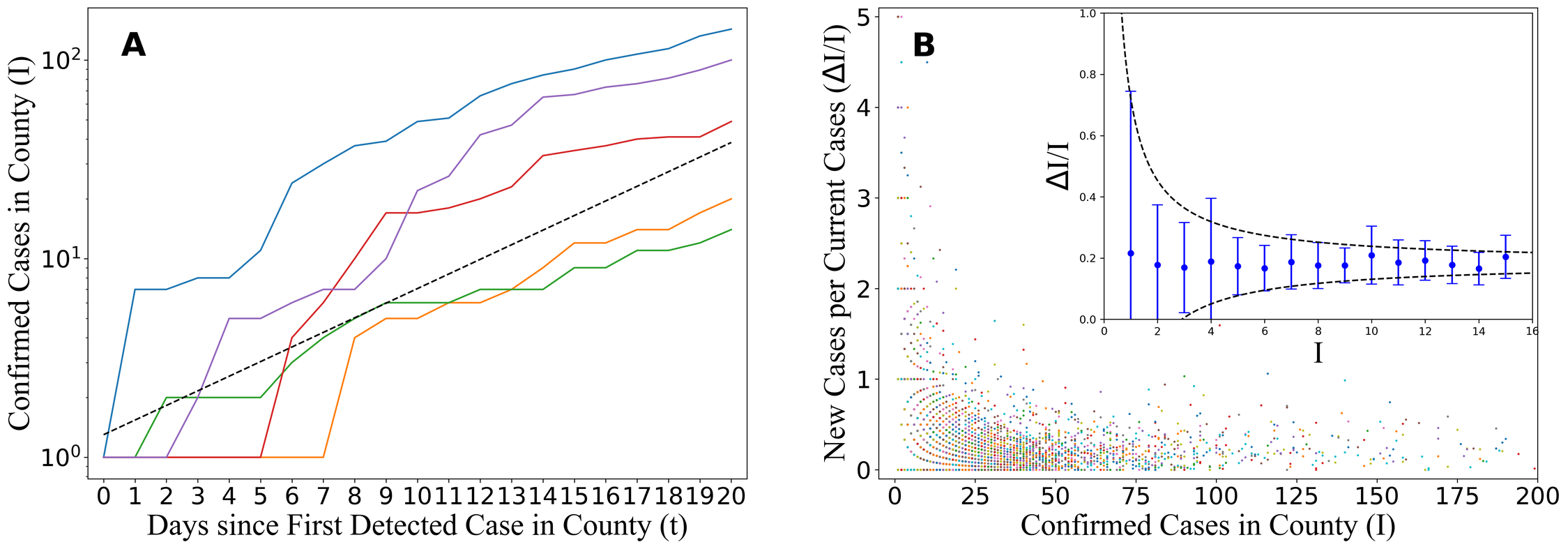}
\caption{\label{growth_rate} (a) Illustration of the variance in early-time growth rate of new cases. At early times, there is noticeable variance in the growth rate between counties. As the number of cases grows, all counties stabilize towards the average growth rate $I \sim (1+\mu_\beta)^{t}$, (dashed black line) where $t$ is the number of days since the first case in a county. The counties shown are Boulder, CO (blue), St. Mary, LA (purple), Vanderburgh, IN (red), Mesa, CO (orange), and Jones, GA (green). (b) The number of daily infections per infected individual as a function of total infections. In the main figure, each point corresponds to a given county (across all US counties that never report $\Delta I<0$) at a given time point (within the first 14 days after the first infection reported in that county). As the number of cases increase, all counties converge to the mean infection rate. The mean (points) and variance (bars) of $\Delta I / I$ at a given $I$ are shown in the inset. The variance decreases like $(\mu_\beta+\sigma_\beta^2)/I$ (black lines).}
\end{figure}

Formalizing this idea, we present a derivation of the variance in the exponential growth rate, or number of new cases per infected individual per day, $\Delta I/ I$, using an SIR framework that incorporates a probability distribution for the infectiousness of a given individual. Our result implies a simple method for estimating the mean, $\mu_\beta$, and variance, $\sigma_\beta^2$, of the infectiousness $\beta$. We apply this method to data for COVID-19 cases in the USA, and find a mean infection rate of $\mu_\beta = 0.18$ cases/day and standard deviation of $\sigma_\beta \gtrsim 0.59$ cases/day. Since the standard deviation is considerably larger than the mean, with $\sigma_\beta/\mu_\beta \gtrsim 3.2$, we conclude that superspreading is prevalent. By our estimate, these results imply that at least $81\%$ of new cases are caused by the top $10\%$ of most infectious individuals. Our method, which uses only a direct measurement of variance in detected case data in the USA, is consistent with estimates of superspreading using surveillance data \cite{georgia}, secondary-case data \cite{hasan2020superspreading}, and more complicated estimates of cluster size distribution using Markov Chain Monte Carlo \cite{overdisp}.

\section*{Results}

\subsection*{\label{SIR_model}Variance in Growth Rate in the SIR Model}
We derive a relation between the variance in the case growth rate and the variance in individual infectiousness between individuals in the population. We start with a standard discrete-time SIR model \cite{SIR}, which is governed by the following difference equations:
\begin{align}
\begin{split}
\Delta S &= -\beta I\frac{S}{N} \\
\Delta I &= \beta I\frac{S}{N} - rI \\
\Delta R &= rI
\end{split}
\end{align}
Here, $N$ is the total population and $S$, $I$, and $R$ are the time-dependent numbers of susceptible, infected, and recovered individuals, respectively. The parameters $\beta$ and $r$ encode the infectiousness and recovery rate of a disease within a population. The time is effectively discretized into days by the available data, so we use $\Delta I$ rather than the usual time derivative, $dI/dt$. The SIR description typically assumes fixed values for $\beta$ and $r$ across the population. However, in superspreading contexts there is a substantial variance in the infectiousness within a population \cite{nature,superspread_past,superspreaders1,superspreadinghist}. We account for this variation by introducing a probability distribution of infectiousness, $p(\beta)$, so that the probability for a randomly-selected individual to have infectiousness in the range $(\beta, \beta+d\beta)$ is given by is given by $p(\beta)d\beta$.

For an individual with a given infectiousness, $\beta$, the probability of infecting exactly $n$ others in a day follows the Poisson distribution, $\textrm{Pois}(n;\beta)$. The probability that a randomly selected individual will infect $n$ others is given by combining the Poisson distribution with the distribution $p(\beta)$, giving
\begin{equation}
P(n) = \int_0^\infty d\beta \frac{e^{-\beta}\beta^n}{n!}p(\beta).
\end{equation}
The first two moments of $P(n)$, $\mu_n$ and $\sigma_n^2$, can be calculated independent of the form of $p(\beta)$:
\begin{align}
\mu_n = \sum_{n=0}^\infty n P(n) &= \mu_\beta
\\
\label{varn}
\sigma_n^2 = \sum_{n=0}^\infty (n-\mu_n)^2 P(n) &= \mu_\beta + \sigma_\beta^2
\end{align}
Equation (\ref{varn}) represents the variance, among all infected individuals, of the number of new infections caused by a single person in a given day. When there are $I$ active cases, the mean number of new cases per infected person, $\Delta (I+R)/I$, is given by the average of $I$ random variables drawn from the distribution $P(n)$. By the central limit theorem, it follows that $\textrm{Var}(\Delta (I+R)/I) = \sigma_n^2/I$. 
Additionally, in the SIR model with a finite total population $N$, $\Delta (I+R)/I = \beta S/N = \beta(1-(I+R)/N)$ decreases as the susceptible population continually shrinks. Effectively, $p(\beta)$ is scaled by the factor $(1-(I+R)/N)$, which represents the fraction of the population that remains susceptible. Consequently, $\mu_\beta \rightarrow \mu_\beta(1-(I+R)/N)$ and $\sigma_\beta^2 \rightarrow \sigma_\beta^2(1-(I+R)/N)^2$. Therefore the total variance in $\Delta (I+R)/I$ follows:
\begin{equation}\label{SIReq}
\text{Var}\left(\frac{\Delta (I+R)}{I}\right) =\frac{\mu_\beta\left(1-\frac{I+R}{N}\right) + \sigma_\beta^2\left(1-\frac{I+R}{N}\right)^2}{I}
\end{equation}
This result becomes simpler in the limiting case where there is no significant change in the susceptible population ($N \rightarrow \infty$) and no recovery ($R \rightarrow 0$). In this limit, we retrieve the case of simple exponential growth, for which \cite{exp_eq}
\begin{equation}\label{exp}
\text{Var}\left(\frac{\Delta I}{I}\right) =\frac{\mu_\beta + \sigma_\beta^2}{I}.
\end{equation}
In the limit $\sigma_\beta \rightarrow 0$, where every infected individual has the same infectiousness $\mu_\beta$, the variance in the average infection rate is simply $\mu_\beta/I$, which corresponds to the variance in a Poisson process with rate $\mu_\beta$. 

In the case of SARS-CoV-2, it is well established that there are asymptomatic carriers \cite{asymp, asymptomatic, asymptomatic2} who transmit the virus without being detected, as well as other infections that are undetected or unreported. Current estimates typically predict that only $10-25\%$ \cite{undetected_early,undet, lu_estimating_2020} of cases are detected. One can attempt to address this effect by assuming that there is a fixed detection probability, $p_\text{det}$, and that the entire infected population, regardless of symptoms, follows the same infectiousness distribution $p(\beta)$. In this case, there are many more infected individuals, $I \sim I_\text{det}/p_\text{det}$, than those detected, which reduces the statistical fluctuations in the growth rate and makes our calculation of $\sigma_\beta^2$ a lower bound. The effect of undetected cases is considered in more detail in \hyperref[S3_App]{Appendix C}. In order to be conservative (especially given the possibility that asymptomatic cases have a lower rate of infection than symptomatic ones \cite{wang_characterization_nodate, chu2020physical}), the results we present here use $p_\text{det} = 1$.

\subsection*{\label{data}Data for COVID-19 in the USA}

We now turn our attention to data for total detected cases of COVID-19 in the USA, taken from the publicly available data set at Ref.~\cite{data}. In the following analysis we limit our consideration to only a short timescale ($\sim$14 days) after the first infection is detected in a given county. This limitation in time scale serves three main purposes; first, it is likely that through changes in policy, lockdown, social distancing, mask usage, etc., the average infectiousness within the population is time-dependent. By restricting ourselves to a relatively small window of early times, we may assume that there is a constant average infectiousness. Second, considering only beginning stages allows us to neglect the possible saturation of the susceptible population, effectively allowing us to take the $N \rightarrow \infty$ limit. Finally, the recovery period for COVID-19 ranges from 7-14 days \cite{recoveryperiod, params} and so by considering this two week period, we can treat our system as if there is limited recovery and $R \rightarrow 0$. These restrictions allow us to treat the USA data using the exponential case, Eq (\ref{exp}).

\begin{figure}[hpt!]
\centering
\includegraphics[scale = 0.5]{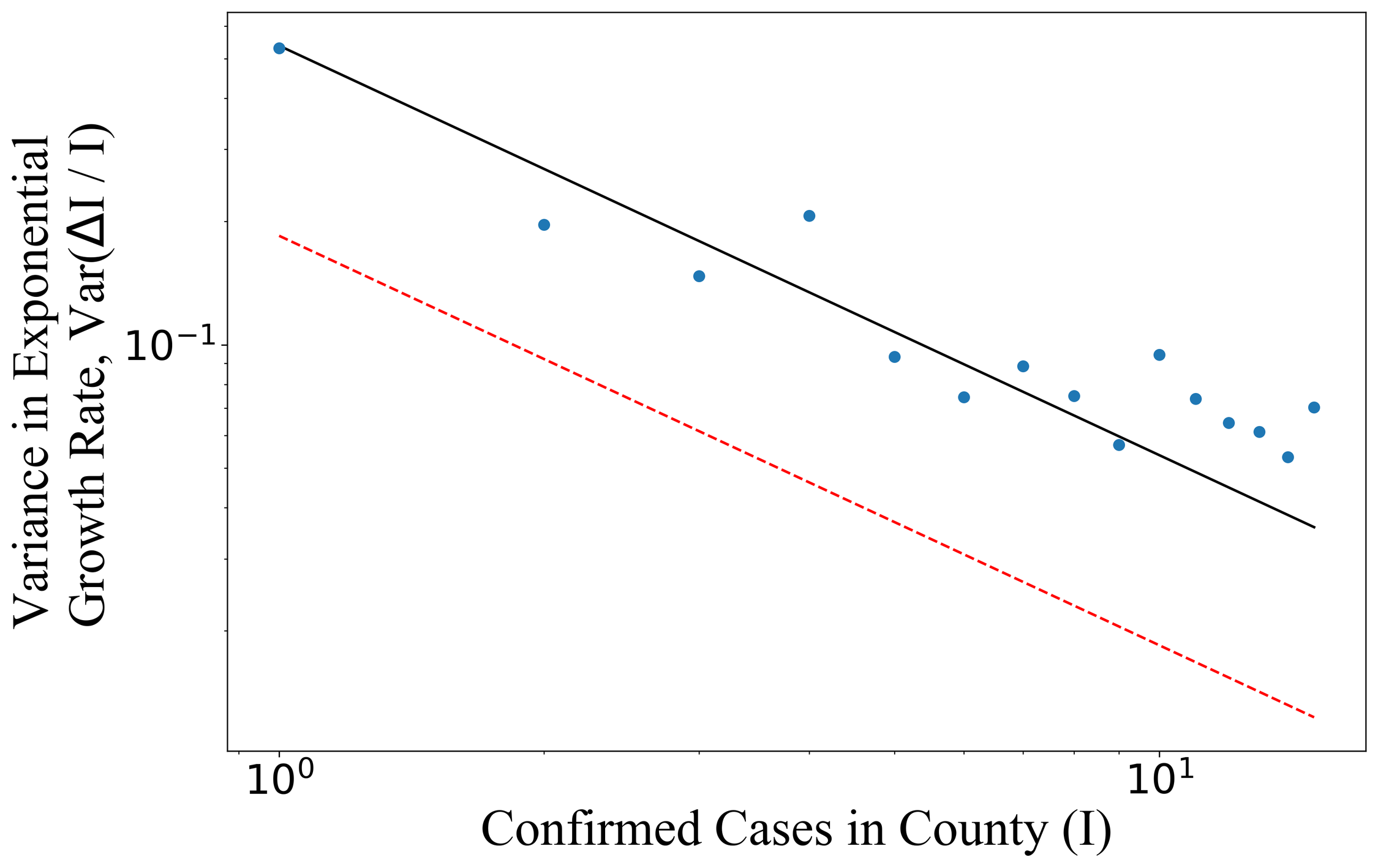}
\caption{As the number of infections $I$ in a given county increases, the variance in exponential growth rate, $\text{Var}(\Delta I / I)$, decreases as $(\mu_\beta+\sigma_\beta^2)/I$. Each data point at a given $I$ is calculated by taking the sample variance in $\Delta I / I$ across all counties when they have $I$ cases. We observe that the USA data (blue) is inconsistent with a model of uniform infectiousness, or $\sigma_\beta = 0$ (dashed red line). A fit to the data (solid black line) implies a large variance in infectiousness, such that $\sigma_\beta/\mu_\beta \gtrsim 3.2$.}
\label{USA_var} 
\end{figure}

In our analysis, the population is divided into geographic regions and the variance is calculated across different trajectories $I(t)$. The US cases are divided by county. For each county, we calculate the average number of new cases per current case per day, $\Delta I/I$, for the first 14 days after the first infection is detected in that county. The variance in $\Delta I/I$ is then calculated among all counties that have a given fixed value of $I$ (we present data only for values of $I$ that have at least 250 corresponding counties). As shown in Fig \ref{USA_var}, the US data generally follows the predicted $\sim 1/I$ trend. An unbiased fit of the data gives $\textrm{Var}(\Delta I/I) \propto I^{-0.74}$. From Eq (\ref{exp}), we calculate $\mu_\beta+\sigma_\beta^2$ by averaging Var$(\Delta I/I) \times I$, weighted by the number of instances at each $I$ value. One might worry that the main source of variation comes from differing average growth rates, $\mu_\beta$, in various counties (e.g.\ rural vs.\ urban). However, we show in the \hyperref[S2_App]{Appendix B} that variance in $\mu_\beta$ across counties is too small to explain the large observed variance in $\Delta I / I$. 

We calculate $\mu_\beta$ from the entire USA population by averaging all values of $\Delta I/I$ weighted by the current number of infections. Equivalently, we sum the number of cases caused each day and then divide by the sum of the number of cases across those days. This procedure gives the mean infectiousness, $\mu_\beta$, and thus from Eq (\ref{exp}) and the fitted slope in Fig \ref{USA_var}, we can infer $\sigma_\beta^2$.

This calculation yields $\mu_\beta = 0.18$ cases/day and $\sigma_\beta = 0.59$ cases/day. The small value of $\mu_\beta^2/\sigma_\beta^2 = 0.096$, equivalent to the dispersion parameter \cite{network_1,overdisp, overdisp_param}, provides clear evidence for superspreading during early stages of the COVID-19 pandemic in the United States. (See \hyperref[S7_App]{Appendix G} for discussion about defining the dispersion parameter in terms of the daily infection rate.)

These results for $\mu_\beta$ and $\sigma_\beta$ can be used to further quantify the extent of superspreading under the assumption that $p(\beta)$ follows a gamma distribution (as in Ref. 18). In the Methods section we present a derivation of the cumulative share of infections, $Y$, caused by the top $X$ portion of most infectious cases. The corresponding ``Lorenz curve'' $Y(X)$ is plotted in Fig 3. This result implies (using our relatively conservative estimate of $\sigma_\beta$) that 81\% of new infections are produced by the top 10\% of most infectious individuals, while only about 4.5\% of cases arise from the 80\% of infected individuals with the lowest infection rates.

\begin{figure}[htp]
\centering
\includegraphics[scale = 0.5]{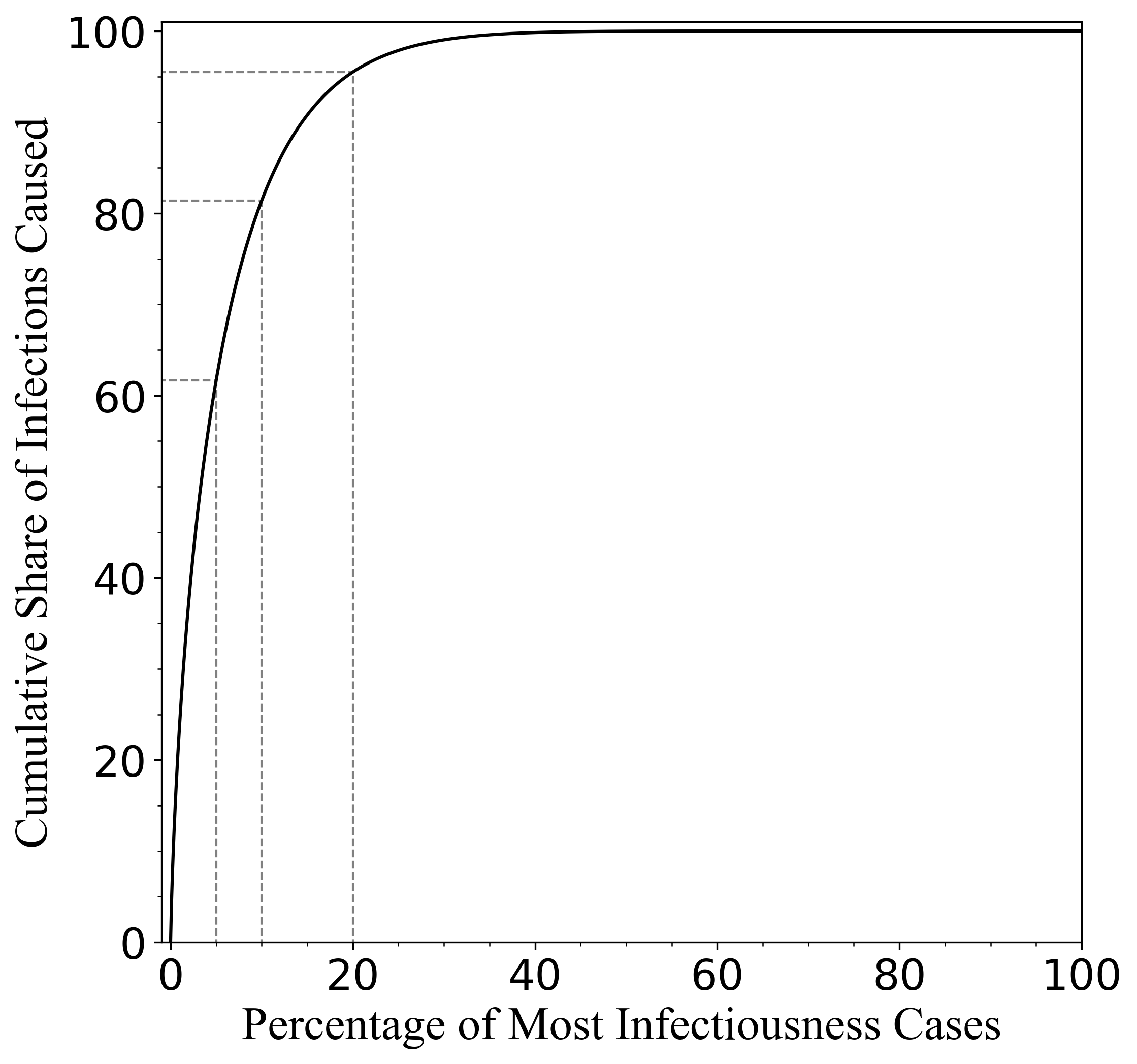}
\caption{An estimated Lorenz curve for SARS-CoV-2 infections in the USA, which displays the percentage of new cases that are caused by a given cumulative percentage of most infectious individuals (solid black). A few points in the curve are highlighted (dashed grey lines): $61.7\%$, $81.4\%$, and $95.5\%$ of new cases are caused by the top $5\%$, $10\%$, and $20\%$ infectious cases, respectively. Accounting for undetected and asymptomatic cases would apparently make this curve steeper, corresponding to more severe superspreading.}
\label{lorenz} 
\end{figure}

\section*{Discussion}
As we have shown, a wide distribution $p(\beta)$ in infectiousness $\beta$ leads to large statistical variation in the early-time growth rate of a disease. By calculating the variance in growth rate among different subpopulations one can infer the variance in $p(\beta)$. Our result for COVID-19 cases in the USA suggests that $\sigma_\beta/\mu_\beta \gtrsim 3.2$, implying a relatively severe superspreading. If we further assume that $p(\beta)$ follows a gamma distribution (as in Ref.\ \cite{nature}), then we can produce a more direct estimate of the extent of superspreading (Fig \ref{lorenz}). Our relatively simple and direct method, based on a calculation of variance in reported case data, can be contrasted with more complicated methods for inferring the dispersion parameter that are based on maximum likelihood estimation (e.g., Ref. \cite{overdisp_param} develops such a method using simulated data), cluster size distributions \cite{overdisp,MERS}, and surveillance or tracing data \cite{georgia, hasan2020superspreading}. These methods also tend to yield a lower-bound estimate for $\sigma_\beta/\mu_\beta$. While studies based on testing and contact tracing (e.g., Refs.~\cite{althaus_ebola_2015, melsew_role_2019, adegboye2018network, nature}) remain the definitive method for assessing superspreading, the method we present here may provide a much simpler way of estimating its prevalence across a much larger population.

We emphasize that our analysis is unable to determine whether this large variance is a result of differing biological symptoms, social behavior, or other possible explanations. Additionally, this estimation is carried out for early times to minimize effects from a time varying $p(\beta)$ and therefore predominantly speaks to the infectiousness prior to widespread lockdown measures.

We close by commenting on a number of complicating factors that we did not include in our analysis and which, one might suspect, could alter our primary finding of a large value of $\sigma_\beta / \mu_\beta$. For example, we have assumed a uniform value of $\mu_\beta$ across different geographic locations; we have neglected undetected cases; we have ignored the possible variation in detection rate $p_\text{det}$ among different counties; we have effectively treated each county as an isolated population and have neglected cross-county interactions; and we have ignored the effects of the incubation period as well as the potential variation in incubation periods between individuals. In the Supplemental Information, we consider each of these mechanisms in turn and show that none of them can explain our result, so that our conclusion of prevalent superspreading of SARS-CoV-2 in the USA remains robust. In brief: the variation in $\mu_\beta$ among different geographic locations is too small to explain the observed variance in growth rate [\hyperref[S2_App]{App. B}]; neglecting undetected cases leads to an \emph{underestimate} of the variance $\sigma_\beta^2$, so that our result is effectively a lower bound for the prevalence of superspreading [\hyperref[S3_App]{App. C}]; variation in $p_\text{det}$ between counties does not directly affect the variance in the growth rate $(\Delta I_\text{det})/I_\text{det}$, other than to provide an average of $p_\text{det} < 1$, which results in a lower-bound estimate of $\sigma_\beta^2$ [\hyperref[S4_App]{App. D}]; cross county interactions tend to reduce the variance, so our result cannot be explained as a consequence of such interactions [\hyperref[S5_App]{App. E}]; and variations in incubation period can only reduce the apparent variance in growth rate [\hyperref[S6_App]{App. F}].

\section*{Methods}

\subsection*{Data source}
We use publicly available data taken from the data set provided by the Center for Systems Science and Engineering (CSSE) at Johns Hopkins University \cite{data} to estimate $\mu_\beta$. Knowing  $\mu_\beta$ enables us to determine $\sigma_\beta$  by taking a best fit to Eq (\ref{exp}). Counties that recorded $\Delta I < 0$ at any point are discarded from the analysis due to the potential for recording error; such counties comprise $\sim$20\% of all counties.

\subsection*{Numerical simulation}
We corroborate Eqs (\ref{SIReq}) and (\ref{exp}) using a numerical simulation of the trajectories of infection growth, $I(t)$, for a given distribution $p(\beta)$. Reference 18 has suggested that infectiousness follows a gamma distribution, and consequently, $P(n)= \textrm{NB}(n; \mu_\beta^2/\sigma_\beta^2,\mu_\beta/(\mu_\beta+\sigma_\beta^2))$ where NB is the negative binomial distribution \cite{reopen,overdisp}. Using this assumption, we simulate the growth of the epidemic by assuming that a given individual $i$, with infectiousness $\beta_i$ that is drawn randomly from $p(\beta)$, generates a number $n_i$ of new cases each subsequent day that is drawn from $\textrm{Pois}(n_i;\beta_i)$. The simulation results confirm Eqs (5) and (6), as shown in \hyperref[S1_App]{Appendix A}. Numerical simulations were performed using Python; the primary analysis is publicly available \cite{code} and the simulations are available upon request to the corresponding author.

\subsection*{Derivation of the curve $Y(X)$}

Following Ref. 18, we assume that the distribution of infectiousness, $p(\beta)$, follows a gamma distribution. This assumption also allows us to further quantify the degree of superspreading by deriving a mathematical relation for the curve $Y(X)$, where $Y$ represents the proportion of infections produced by the top $X$ fraction of most infectious individuals.  In particular, one can calculate the fraction of individuals $X_{\beta_0}$ with infectiousness larger than a given value $\beta_0$, as well as the fraction of secondary infections $Y_{\beta_0}$ that these individuals are expected to cause:

\begin{equation}
X_{\beta_0} = \int_{\beta_0}^{\infty}d\beta \, p(\beta) = Q \left( \frac{\mu_\beta^2}{\sigma_\beta^2}, \beta_0 \frac{\mu_\beta}{\sigma_\beta^2} \right)
\end{equation}
\begin{equation}
    Y_{\beta_0} = \int_{\beta_0}^{\infty}d\beta \, p(\beta)\frac{\beta}{\mu_\beta} = Q \left( 1+\frac{\mu_\beta^2}{\sigma_\beta^2}, \beta_0 \frac{\mu_\beta}{\sigma_\beta^2} \right),
\end{equation}
where $Q$ is the Regularized Gamma function. By eliminating $\beta_0$ we find
\begin{equation}\label{lorenzeq}
Y = Q\left(1+\frac{\mu_\beta^2}{\sigma_\beta^2},Q^{-1}\left(\frac{\mu_\beta^2}{\sigma_\beta^2}, X\right)\right).
\end{equation}

Figure \ref{lorenz} displays the cumulative share of infections, $Y$, caused by the top $X$ portion of most infectious cases.

\section*{Acknowledgments}
The authors are grateful to N.\ E.\ Skinner for helpful conversations.

\providecommand{\noopsort}[1]{}\providecommand{\singleletter}[1]{#1}%

\makeatletter

\renewcommand\refname{References A}
\renewcommand\@bibitem[1]{\item\if@filesw \immediate\write\@auxout
    {\string\bibcite{#1}{A\the\value{\@listctr}}}\fi\ignorespaces}
\def\@biblabel#1{[A#1]}
\makeatother

\section*{\label{S1_App}Appendix A: Simulations.}

We employ Monte Carlo simulations to corroborate our theoretical calculations [Eqs (5) and (6)]. We start by adopting the conclusion of Ref.\ \cite{natureA} and defining $p(\beta)$ as a Gamma distribution with mean $\mu_\beta$ and standard deviation $\sigma_\beta$. We simulate an outbreak by first randomly generating a $\beta$ from the distribution $p(\beta)$ and then drawing a random $n$ from $\textrm{Pois}(n;\beta)$. This process is repeated until a non-zero $n$ is generated, representing an outbreak starting in a given location with $n$ individuals. Each of these $n$ infected individuals is given a correspond infectiousness, $\beta_i$, which they keep for the remainder of the simulation. Each individual, $i$, then generates $n_i$ new cases randomly drawn from $\textrm{Pois}(n_i;\beta_i)$, and each of these secondary infections is assigned its own randomly generated infectiousness as well. During every iteration of the simulation, representing a day, each infected individual infects others given by a new random Poisson variable with mean defined by their own infectiousness. After a set number of infections is reached, the simulation is stopped and the trajectory $I(t)$ is recorded. This process is repeated for 3,000 total trajectories, representing the $\sim$3,000 counties in the real USA data. This simulated data is then treated in the same manner as the real data, which is explained in ``Data for COVID-19 in the USA" of the main text.

\begin{figure}[hbt!]
\centering
\includegraphics[scale = .75]{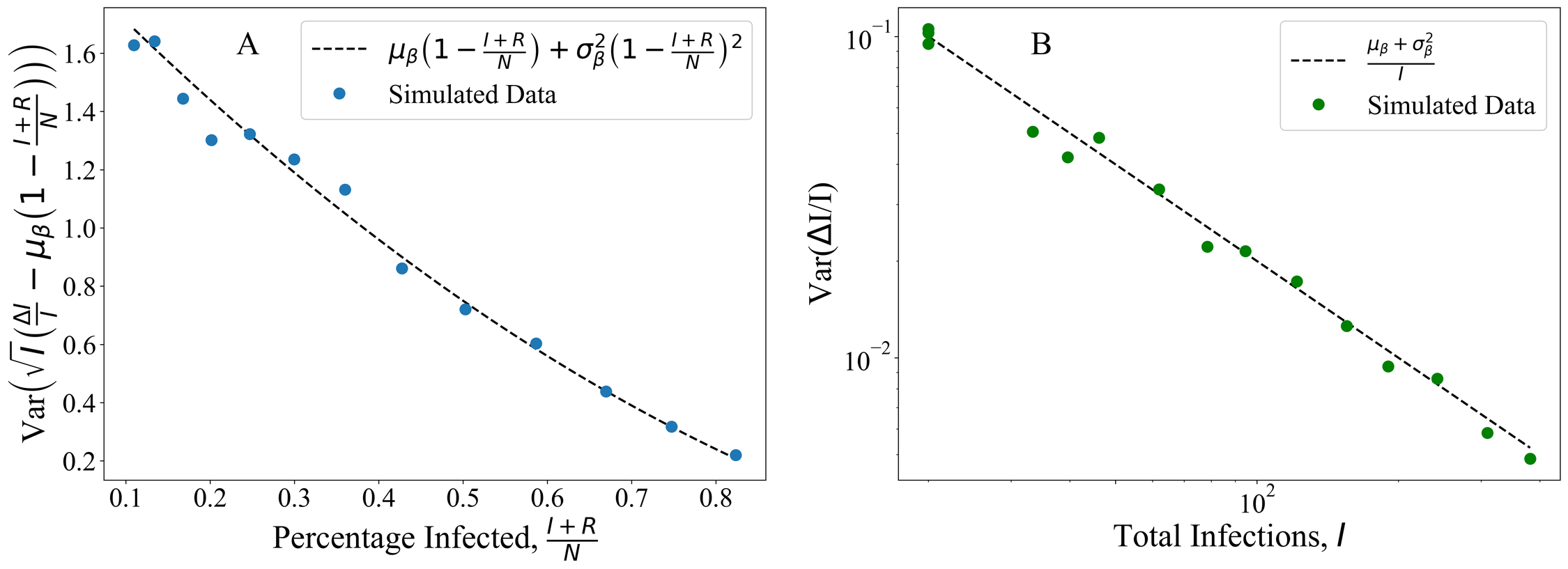}
\caption{\label{SIR_vs_Exp} (a) Variance in the scaled growth rate for a simulated SIR model with various $N$ values. The variance decreases as the susceptible population diminishes and infected individuals recover. (b) Simulation of an exponential model with no recovery, which is equivalent to early times in the pandemic. The variance in average infection rate starts at $\mu_\beta + \sigma_\beta^2$ at $I = 1$ and then decreases as $\sim 1/I$ as the number of infected individuals increases.}
\end{figure}

We also consider simulations with a recovery phase and a finite carrying capacity $N$. To implement recovery, we specify a given number of days, $t_\text{rec}$, over which an infected individual is infectious. Only those who contract the virus within this time period infect others. The effect of finite carrying capacity $N$ is included by scaling the infectiousness $\beta_i$ of individual $i$ by the factor $S/N = 1 - (I+R)/N$. For example, someone with infectiousness $\beta_i$ generates a number of cases $n_i$ drawn from the probability distribution $\textrm{Pois}(n;\beta_i (1- (I+R)/N))$ each day. This procedure is cut off once a certain fraction of $N$ is reached, and then repeated 3,000 times. Although each trajectory follows the same $p(\beta)$, $N$ can vary between different trajectories. To account for this variation in $N$, we normalize $\Delta I \rightarrow \sqrt{I}(\Delta I/I - \mu_\beta (1-(I+R)/N))$. We see that the simulated variance matches well with our theory (Fig \ref{SIR_vs_Exp}).

\section*{\label{S2_App}Appendix B: Variance in $\mu_\beta$.}

\begin{figure}[hbt!]
    \centering
    \includegraphics[scale = 0.75]{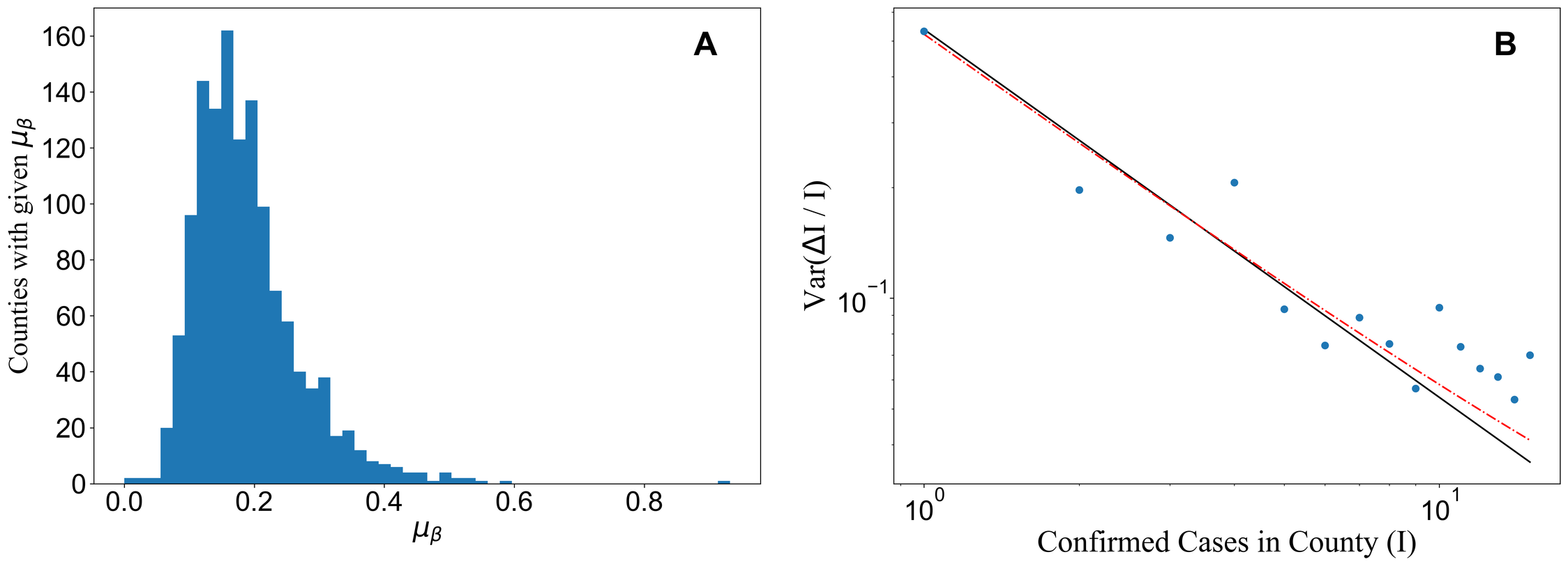}
    \caption{(a) The calculated value of the mean infectiousness, $\mu_\beta$, for each individual county (with at least five cases). The variance in $\mu_\beta$ is relatively small: Var$(\mu_\beta) = 0.0068$ (cases/day)$^2$ $\ll$ $\sigma_\beta^2$. (b) When we account for this consideration (dashed red line), the fitted value of $\sigma_\beta^2$ decreases from $0.35$ $\rightarrow$ $0.33$ (cases/day)$^2$. This adjustment does not significantly affect our conclusions.}
    \label{var_muk}
\end{figure}

It is reasonable to question whether the calculated variance, $\sigma_\beta^2$, is a result of various geographic locations having differing average infectiousness, $\mu_\beta$, due to varying population density, social norms, etc. One may instead consider that the mean infectiousness $\mu_\beta$ follows some distribution $q(\mu_\beta)$ among different counties. For a given $\mu_\beta$, we have shown that the variance in $P(n;\mu_\beta,\sigma_\beta)$ averaged over $I$ realizations is given by $\left(\mu_\beta^2 + \sigma_\beta^2\right)/I$ in the exponential case. Including the effect of a distribution $q(\mu_\beta)$, we calculate the variance in $\Delta I/I$ to be: 

\begin{align*}
    \text{Var}\left(\frac{\Delta I}{I}\right) &= \sum_{n=0}^{\infty} (n-\bar{\mu}_\beta)^2 \int_0^{\infty} d\mu_\beta \, q(\mu_\beta)P(n;\mu_\beta,\sigma_\beta)\\
    &= \int_0^{\infty} d\mu_\beta \, q(\mu_\beta)\sum_{n=0}^{\infty} (n-\bar{\mu}_\beta)^2 P(n;\mu_\beta,\sigma_\beta)\\
    &=\int_0^{\infty} d\mu_\beta \, q(\mu_\beta) \left((\mu_\beta-\bar{\mu}_\beta)^2 +\frac{\mu_\beta+\sigma_\beta^2}{I}\right)\\
    &= \text{Var}(q(\mu_\beta)) + \frac{\bar{\mu}_\beta+\sigma_\beta^2}{I}
\end{align*}
That is, when we account for the possibility that each region has a different $\mu_\beta$, the value of $\mu_\beta$ is replaced by its mean $\bar{\mu}_\beta$ across counties, and a constant term is added for the variance in $\mu_\beta$ across counties. We can conclude that this variance cannot fully explain the data for two reasons. First, we observe a clear Var$(\Delta I/I)\sim 1/I$ trend in the data (Fig \ref{var_muk}), which can only be a result of the variance in $p(\beta)$ rather than $q(\mu_\beta)$. Additionally, we can directly measure the variance in $\mu_\beta$ across counties, which we find to be $0.007$ (cases/day)$^2$. This number is too small to significantly affect the total variance in $\Delta I/I$, as seen in Fig \ref{var_muk}. When the measured variance in $q(\mu_\beta)$ is taken into account in our fitting procedure, we find that $\bar{\mu}_\beta = 0.18$ cases/day, $\sigma_\beta \gtrsim 0.58$ cases$^2$/days$^2$, resulting in very slightly different value of $\sigma_\beta/\mu_\beta \gtrsim 3.1$. 

\makeatletter

\renewcommand\refname{References C}
\renewcommand\@bibitem[1]{\item\if@filesw \immediate\write\@auxout
    {\string\bibcite{#1}{C\the\value{\@listctr}}}\fi\ignorespaces}
\def\@biblabel#1{[C#1]}
\makeatother

\section*{\label{S3_App}Appendix C: Undetected cases.}
Between asymptomatic cases and imperfect testing, there are a significant number of active cases, which can transmit the virus, that do not show up in the data set we use. One way to take this effect into account is by introducing an average probability that a case is detected, $p_\text{det}$. All variance calculations occur at fixed values of the number of detected cases, $I_\text{det}$. Given $I_\text{det}$, the probability that there are $I$ total cases is given by a negative binomial distribution:
\begin{equation}
    P(I;I_\text{det}) = {I-1 \choose I_\text{det}-1}{p_\text{det}}^{I_\text{det}}(1-p_\text{det})^{I - I_\text{det}}.
\end{equation}
If there are $I$ active cases, then the probability that $\Delta I$ cases are generated is given by the sum of $I$ random variables drawn from $P(n)$. Since $\mu_n = \mu_\beta$ and $\sigma_n^2 = \mu_\beta+\sigma_\beta^2$,
\begin{equation}
    \sum_{\Delta I = 0}^{\infty} P(\Delta I;I)\Delta I = \mu_\beta I
\end{equation}
\begin{equation}
\sum_{\Delta I = 0}^{\infty} P(\Delta I;I)(\Delta I-\mu_\beta I)^2  = (\mu_\beta+\sigma_\beta^2) I.
\end{equation}
Once $\Delta I$ cases are generated on a given day, the probability that $\Delta I_\text{det}$ are detected is given by a binomial distribution: 
\begin{equation}
    \label{PDeltaI}
    P(\Delta I_\text{det};\Delta I) = {\Delta I \choose \Delta I_\text{det}}{p_\text{det}}^{\Delta I_\text{det}}(1-p_\text{det})^{\Delta I - \Delta I_\text{det}}.
\end{equation}

We combine these equations to derive the probability distribution for the number of new detected cases in a given day, $\Delta I_\text{det}$, given that there are currently $I_\text{det}$ active cases.
\begin{equation}
P(\Delta I_\text{det}; I_\text{det})= \sum_{I=I_\text{det}}^{\infty} P(I; I_\text{det}) \sum_{\Delta I=\Delta I_\text{det}}^{\infty} P(\Delta I;I) P(\Delta I_\text{det}; \Delta I).
\end{equation}
It follows that the mean and variance in $\Delta I_\text{det}/I_\text{det}$ are
\begin{align}
\begin{split}
    \textrm{Mean}\left(\frac{\Delta I_\text{det}}{I_\text{det}}\right) &= \frac{1}{I_\text{det}}\sum_{\Delta I_\text{det} = 0}^{\infty}P(\Delta I_\text{det}; I_\text{det}) \Delta I_\text{det}\\
    &= \mu_\beta
\end{split}
\end{align}
\begin{align}
\begin{split}
    \label{finalvar}
    \textrm{Var}\left(\frac{\Delta I_\text{det}}{I_\text{det}}\right) &= \frac{1}{I_\text{det}^2}\sum_{\Delta I_\text{det} = 0}^{\infty}P(\Delta I_\text{det}; I_\text{det}) (\Delta I_\text{det}- I_\text{det}\mu_\beta)^2\\
    &= \frac{\mu_\beta+\mu_\beta^2(1-p_\text{det})+p_\text{det}\sigma_\beta^2}{I_\text{det}}.
\end{split}
\end{align}
That is, when under-detection is accounted for, an extra term $\mu_\beta^2(1-p_\text{det})/I_\text{det}$ is added to the variance due to the variance in the underlying total number of cases, $I$. The term $\sigma_\beta^2/I_\text{det}$ is also scaled down by a factor $p_\text{det}$, since there are on average a larger number $I\sim I_\text{det}/p_\text{det}$ of total cases. Thus, since $\mu_\beta^2(1-p_\text{det})\ll \mu_\beta$ and $\sigma_\beta^2$ is suppressed by a factor of $p_\text{det}$, the calculated value of $\sigma_\beta^2$ is a lower bound.

Although there are estimates of the fraction of detected COVID-19 cases in literature (e.g., Refs.~\cite{undetected_earlyC,undetC, lu_estimating_2020C}), in order to be conservative we do not directly use these estimates for the parameter $p_\text{det}$ in Eq (\ref{finalvar}), due to the possibility that undetected cases have a different (likely lower) infectiousness.

\providecommand{\noopsort}[1]{}\providecommand{\singleletter}[1]{#1}%

\makeatletter
\renewcommand\refname{References D}
\renewcommand\@bibitem[1]{\item\if@filesw \immediate\write\@auxout
    {\string\bibcite{#1}{D\the\value{\@listctr}}}\fi\ignorespaces}
\def\@biblabel#1{[D#1]}
\makeatother

\section*{\label{S4_App}Appendix D: Variance in testing.}
We have shown that a fixed detection rate, $p_\text{det}$, across counties cannot account for the variance observed within the US population. However, one can also check to ensure that variation in $p_\text{det}$ between counties, described by a probability distribution $q(p_\text{det})$, does not explain the data either. To account for differing values of $p_\text{det}$ we weight Eq (\ref{PDeltaI}) by $q(p_\text{det})$ so that $P(\Delta I_\text{det};\Delta I) \rightarrow \int_{0}^{1} dp_\text{det} q(p_\text{det})P(\Delta I_\text{det};\Delta I)$. Plugging into Eg.~(\ref{finalvar}) we see
\begin{equation}
\label{pdet}
\textrm{Var}\left(\frac{\Delta I_\text{det}}{I_\text{det}}\right) = \frac{\mu_\beta+\mu_\beta^2(1-\overline{p_\text{det}})+\overline{p_\text{det}}\sigma_\beta^2}{I_\text{det}}
\end{equation}
where $\overline{p_\text{det}}$ is the mean detection rate across all counties when they have $I_{\text{det}}$ cases. This expression shows that the variance in the exponential growth rate $(\Delta I_{\text{det}})/I_{\text{det}}$ only depends on the mean detection rate at a given $I_{\text{det}}$ rather than its variance. Furthermore, averaging $p_{\text{det}}$ across counties at various times (but the same $I_{\text{det}}$) will average out any effects from cyclical weekly reporting patterns. To observe how this impacts our calculation for $\sigma_{\beta}^2$, we rearrange Eq. (\ref{pdet}) to obtain:
\begin{equation}
\label{pdet2}
\sigma_\beta^2 = \frac{\textrm{Var}\left(\frac{\Delta I_\text{det}}{I_\text{det}}\right) I_\text{det}-\mu_\beta-\mu_\beta^2(1-\overline{p_\text{det}})}{\overline{p_\text{det}}}
 = \frac{\sigma_{\beta,p_{\text{det}}=1}^2-\mu_\beta^2(1-\overline{p_\text{det}})}{\overline{p_\text{det}}}
\end{equation}
where $\sigma_{\beta,p_{\text{det}}=1}^2$ is the variance we calculate in the main text assuming $\overline{p_\text{det}} = 1$. Since $\mu_\beta^2(1-\overline{p_\text{det}}) \ll \sigma_{\beta,p_{\text{det}}=1}^2$, it is clear that accounting for an imperfect detection rate can only increase the variance in infectiousness.
Therefore, if $\overline{p_\text{det}} < 1$, this makes our calculation a lower bound on $\sigma_\beta^2$. Further, if we use the percentage of asymptomatic cases, 40\% \cite{asymptD}, as a rough estimate for the mean percentage of undetected cases, then $\mu_\beta$ remains unchanged while $\sigma_\beta$ increases from 0.59 cases/day to 0.75 cases/day. This change in the variance corresponds to a significant increase in superspreading as the percentage of new infections cause by the top 5\% of infectious cases rises from 61.7\% to 74.0\%. While this exercise provides some insight into how large $\sigma_\beta$ could be, it is not a rigorous upper bound. Firstly, there remains significant uncertainty in the percentage of asymptomatic cases as estimates range from 8.2\% to 75\% \cite{asympt2D}. Additionally, there remain other complications, such as incubation period variation and cross-county interactions, which would increase the variance further.

\providecommand{\noopsort}[1]{}\providecommand{\singleletter}[1]{#1}%

\section*{\label{S5_App}Appendix E: Cross-county interactions.}
Our analysis in the main text operates under the assumption that each county in the USA is an independent population in which the virus can spread. However, it is clear that there is some portion of infections that cross county lines. To understand how this interaction can affect the variance in observed growth rate of cases, we explore what the variance looks like if we have perfect mixing between $M$ counties, each with $I_1$ active cases. In this formulation, the variance we calculate is $\textrm{Var}(\Delta I_1 /I_1)$ at a given $I_1$. Focusing on a single county with $I_1$ active cases, the probability that there are $I$ total cases across the $M$ counties is given by a negative binomial distribution:
\begin{equation}
P(I;I_1) = {I - 1 \choose I_1 - 1}\left(\frac{1}{M}\right)^{I_1}\left(1-\frac{1}{M}\right)^{I-I_1}.
\end{equation}
We have shown that if there are $I$ total cases, then the mean and variance in the number of new cases, $\Delta I$, is $\mu_\beta I$ and $(\mu_\beta+\sigma_\beta^2)I$, respectively. Once $\Delta I$ cases are generated, they are randomly sorted into the $M$ counties. Therefore, 
\begin{equation}
P(\Delta I_1;\Delta I) = {\Delta I \choose \Delta I_1}\left(\frac{1}{M}\right)^{\Delta I_1}\left(1-\frac{1}{M}\right)^{\Delta I - \Delta I_1}.
\end{equation}
Combining these distributions, the probability that $\Delta I_1$ cases occur in a given county that has $I_1$ active cases is:
\begin{equation}
P(\Delta I_1; I_1)= \sum_{I=I_1}^{\infty} P(I; I_1) \sum_{\Delta I=\Delta I_1}^{\infty} P(\Delta I;I) P(\Delta I_1; \Delta I).
\end{equation}
From $P(\Delta I_{1}; I_{1})$ we calculate the mean and variance in $\Delta I_1 / I_1$ to find:
\begin{align}
\begin{split}
    \textrm{Mean}\left(\frac{\Delta I_1}{I_1}\right) &= \frac{1}{I_1}\sum_{\Delta I_1 = 0}^{\infty}P(\Delta I_1; I_1) \Delta I_1\\
    &= \mu_\beta
\end{split}
\end{align}
\begin{align}
\begin{split}
    \textrm{Var}\left(\frac{\Delta I_1}{I_1}\right) &= \frac{1}{I_1^2}\sum_{\Delta I_1 = 0}^{\infty}P(\Delta I_1; I_1) (\Delta I_1- I_1\mu_\beta)^2\\
    &= \frac{\mu_\beta+\mu_\beta^2\left(1-\frac{1}{M}\right)+\frac{\sigma_\beta^2}{M}}{I_1}.
\end{split}
\end{align}
Since $\mu_\beta = 0.18$ cases/day, the term $\mu_\beta + \mu_\beta^2(1-1/M)$ cannot account for the variation present in the US. Therefore, if there is maximal interactions between counties, then the calculation of $\sigma_\beta^2$ remains a lower bound estimate. Intuitively, one can say that when different counties interact strongly with each other, there is a larger underlying number of active cases from which new cases can be drawn for a given county, and this larger number reduces the statistical variance.

The previous consideration assumes that all counties interact evenly with each other. It is possible that some counties might gain a large number of cases entering from a neighboring county, while others have more exiting than entering. To account for this potential source of variance we consider a single county. Assume there is a fixed portion, $p_\text{exit}$, of new infections from cases within a county that are spread to an outside county, as well as a portion, $p_\text{enter}$, of new cases from other counties. Both of these quantities are defined in terms of the number of cases in the current county, $I(t)$. Consequently, the effective number of cases leading to new infections in the current county is $(1-p_\text{exit}+p_\text{enter})I(t)$. Since all counties are assumed to follow the same underlying distribution $p(\beta)$, we expect that the mean of $\Delta I/I$ will be {$(1-p_\text{exit}+p_\text{enter})\mu_\beta$}. With this understanding, there are two possibilities: either all counties have equal flows in and out so that $p_\text{exit} \sim p_\text{enter}$, or there is a balance between counties with $p_\text{exit} > p_\text{enter}$ and vice versa. In the first case, we see that all counties share approximately the same measured $\mu_\beta$ while in the later, there is a wide spread in $\mu_\beta$ from this cross county interaction. Since we show in \hyperref[S2_app]{Appendix B} that the there is little variance in $\mu_\beta$ across counties, this suggests that $p_\text{exit} \sim p_\text{enter}$ within each county. Consequently, the effective number of cases that can lead to a new case within a given county, $(1-p_\text{exit}+p_\text{enter})I(t)$, varies from the true number, $I(t)$, only on the order of $\sigma_{\mu_\beta} \ll \sigma_\beta$. Therefore, we conclude that this effect cannot explain the large variance we observe.

\section*{\label{S6_App}Appendix F: Variance in incubation period.}
\begin{figure}[hbt!]
    \centering
    \includegraphics[scale = 0.7]{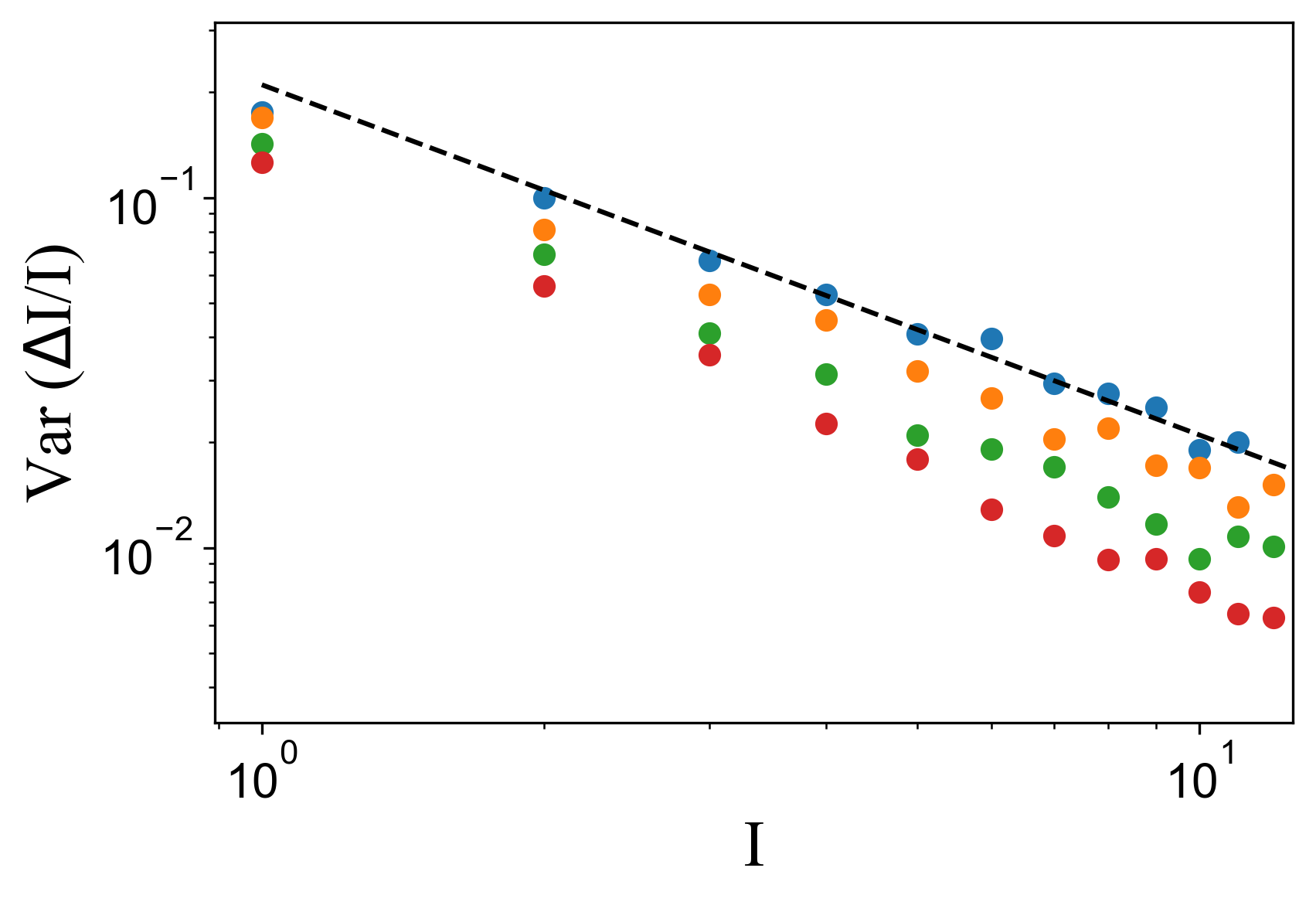}
    \caption{We simulate the impact of variance in incubation time compared to a fixed incubation period (blue). Here, we use $p(L)\sim e^{-\lambda L}$ with $\lambda = 1,2,3$ (orange, green, red). Variance in incubation time only decreases the observed variance and consequently cannot explain the large $\sigma_\beta$ we calculate.}
    \label{latency}
\end{figure}

Our analysis ignores the effect of latency between the moment of infection and the appearance of symptoms that can lead to detection, often called the incubation period. This treatment is equivalent to assuming that there is a fixed incubation period for all infections. In this situation, if the true number of cases at a given time (counting cases as those infected, and not necessarily already detected) is $I(t)$, then the number of cases observed is simply $I_\text{observed}(t) = I(t-L)$, where $L$ is the amount of time for an individual to start showing symptoms and be detected. However, it is reasonable to suspect that some individuals could have shorter or longer incubation periods, and this variance could impact the overall variance that we calculate. To address this, we assume that the duration of incubation time, $L$, follows some distribution $q(L)$ with mean $\mu_L$. At a given time $t$, the infections being observed are originating from times in the vicinity of $t-\mu_L$. Therefore, the effective number of infected individuals leading to new infections at time $t$ is $\sum_{L=\mu_L-\Delta L}^{\mu_L+\Delta L} q(L) I(t-L)$, where $2 \Delta L$ represents the range of possible incubation periods. Since $I(t)$ grows exponentially, the terms from the most recent time with smaller incubation period, $L = \mu_L - \Delta L$, will dominate the sum. Consequently, the effective infected population is always greater than or equal to $I(t-\mu_L)$, and this effect tends to decreases the overall variance in $\Delta I / I$. This diminishing of the variance with increasing $\Delta L$ is corroborated by simulations (Fig \ref{latency}).

\makeatletter
\renewcommand\refname{References G}
\renewcommand\@bibitem[1]{\item\if@filesw \immediate\write\@auxout
    {\string\bibcite{#1}{G\the\value{\@listctr}}}\fi\ignorespaces}
\def\@biblabel#1{[G#1]}
\makeatother

\section*{\label{S7_App}Appendix G: Dispersion parameter comparison.}

The dispersion parameter is typically defined as $\mu_{R_0}^2/\sigma_{R_0}^2$, where $\mu_{R_0}$ and $\sigma_{R_0}^2$ are the mean and variance of the number of new infections caused by an individual over their entire infectious period. Since we work with the daily infection rate, $\beta$, and calculate $\mu_{\beta}^2/\sigma_{\beta}^2$ it is important to draw a direct comparison between these two dispersion parameters.
  
We can understand the distinction between $\mu_\beta^2/\sigma_\beta^2$ and $\mu_{R_0}^2/\sigma_{R_0}^2$ using the SIR-based framework that we establish in the main text. This approach involves calculating the mean ($\mu_{R_0}$) and variance ($\sigma_{R_0}^2$) of $R_0$, the number of infections caused by an infected individual over the course of their infection.

First, if we assume a fixed infection period of $D$ days, then we see that the total number of infections from an individual with infectiousness $\beta$ is the sum of $D$ draws form a Poisson distribution with mean $\beta$. Therefore, 

$$p(R_0;\beta) = \frac{e^{-D\beta}(D\beta)^{R_0}}{R_0!}.$$

Thus across the entire population, with infectiousness described by an arbitrary distribution $p(\beta)$, the distribution of $R_0$ is simply $p(R_0)= \int_0^\infty d\beta p(\beta) p(R_0;\beta)$. From this distribution we can calculate the mean and variance of $R_0$:
\begin{align*}
    \mu_{R_0} &= \sum_{R_0 = 0}^\infty p(R_0) R_0\\
    &= \int_0^\infty d\beta p(\beta) e^{-D\beta}  \sum_{R_0 = 0}^\infty \frac{(D\beta)^{R_0}}{R_0!} R_0\\
    &= \int_0^\infty d\beta p(\beta) D \beta \\
    \mu_{R_0}&= D \mu_\beta
\end{align*}
\begin{align*}
    \sigma_{R_0}^2 &= \sum_{R_0 = 0}^\infty p(R_0) (R_0-\mu_{R_0})^2\\
    &= \int_0^\infty d\beta p(\beta) e^{-D\beta}  \sum_{R_0 = 0}^\infty \frac{(D\beta)^{R_0}}{R_0!} (R_0-D\mu_{\beta})^2\\
    &= \int_0^\infty d\beta p(\beta) \left[D^2(\beta-\mu_\beta)^2 + D\beta\right] \\
    \sigma_{R_0}^2&= D^2\sigma_\beta^2 + D\mu_\beta
\end{align*}

Combining these results we see that the usual dispersion parameter is given by

$$\frac{\mu_{R_0}^2}{\sigma_{R_0}^2} = \frac{ \mu_\beta^2}{\sigma_\beta^2 + \frac{\mu_\beta}{D}}.$$

From this expression, it becomes apparent that while $\mu_{R_0}^2/\sigma_{R_0}^2$ and $\mu_\beta^2/\sigma_\beta^2$ are not exactly equivalent, in our case they are very nearly equal since $\sigma_\beta^2 \gg \mu_\beta/D$. If we let $D = 14$ days and calculate the duration of infection dispersion parameter we obtain $\mu_{R_0}^2/\sigma_{R_0}^2 = 0.093$, as compared to our value of $0.096$.

One can be more careful and let the infectious period be described by an arbitrary distribution $p(D)$ with mean $\mu_D$ and variance $\sigma_D^2$. With this setup, the duration of infection dispersion gains an extra term in the denominator: 

$$\frac{\mu_{R_0}^2}{\sigma_{R_0}^2} = \frac{\mu_{\beta}^2}{\sigma_{\beta}^2+\left(\frac{\sigma_D}{\mu_D}\right)^2(\mu_{\beta}^2+\sigma_{\beta}^2)+\frac{\mu_{\beta}}{\mu_D}}.$$

If we use $\mu_D = 13.4$ days \cite{Byrnee039856G} and conservatively estimate $\sigma_D \sim 5$ days, then the dispersion parameters remain almost unchanged. 

It is worth emphasizing that the primary result of our analysis, the Lorenz curve relationship (Eq.~(9) and Fig.~3 of the main text), does not depend on the difference between these two definitions of the dispersion parameter. The Lorenz curve relation is based on the ratio $\mu_\beta^2/\sigma_\beta^2$, and allows one to obtain results for the percentage of cases caused by (say) the 20$\%$ of most infectious cases. Our results for the Lorenz curve are consistent with other studies, while providing more quantitative detail. 

\providecommand{\noopsort}[1]{}\providecommand{\singleletter}[1]{#1}%

\end{document}